\documentclass[aps,prl,twocolumn,superscriptaddress,preprintnumbers,longbibliography, amsmath,amssymb,10pt,nobibnotes]{revtex4-2}
\usepackage{paperpreamble}

\begin{document}

\title{Extreme fragmentation of a Bose gas}

\author{Nathan Dupont}
\thanks{These authors contributed equally to this work}
\author{Amit Vashisht}
\thanks{These authors contributed equally to this work}
\affiliation{CENOLI,
Universit\'e Libre de Bruxelles, CP 231, Campus Plaine, B-1050 Brussels, Belgium}
\author{Nathan Goldman}
\email{nathan.goldman@ulb.be}
\affiliation{CENOLI,
Universit\'e Libre de Bruxelles, CP 231, Campus Plaine, B-1050 Brussels, Belgium}
\affiliation{Laboratoire Kastler Brossel, Coll\`ege de France, CNRS, ENS-PSL University, Sorbonne Universit\'e, 11 Place Marcelin Berthelot, 75005 Paris, France
}

\date{\today}

\begin{abstract}
    Fragmentation of an interacting Bose gas refers to the macroscopic occupation of a finite set of single-particle eigenstates. This phenomenon is related to the notion of particle-number squeezing in quantum optics, an exquisite property of quantum states that can offer metrological gain.
    So far, fragmentation has  only been partially achieved in experiments involving a large number $N$ of bosons in few modes. Here, we introduce a practical and efficient scheme to prepare fragmented states in systems realizing the $L$-mode Bose-Hubbard model. 
    We demonstrate how a large energy detuning between the modes can be used as a practical control parameter to successfully fragment a Bose gas over an extremely short preparation time. Applying an optimal-control approach within realistic experimental constraints, we obtain total fragmentation at a high filling factor, realizing $\ket{N/L,...,N/L}$ Fock states with hundreds of bosons in very few modes over a few tunneling times.
\end{abstract}

\maketitle

Bose-Einstein condensation (BEC) refers to the macroscopic bosonic occupation of a single-particle ground state~\cite{cornell_2002,ketterle_2002,bloch_2008}. Interestingly, when the single-particle ground state is degenerate, a rich competition occurs between the different states in which to condense.
In such a scenario, if the order of degeneracy $D$ is much smaller than the total number of particles $N$, the system can form a \textit{fragmented} BEC~\cite{nozieres1982,spekkens_1999,mueller_2006} which is characterized by a macroscopic occupation $N/D$ of the degenerate states. 

Ground state degeneracy, stemming from the single-particle Hamiltonian's invariance under a symmetry transformation, is generally lifted, in practice, by arbitrarily small perturbations (e.g.~an external magnetic field) which break the relevant symmetries~\cite{beekman2019}. In practice, this strongly complicates the experimental realization of fragmented BECs with a large $N/D$ ratio. Nonetheless, strong interactions among the particles, with an energy scale larger than symmetry-breaking perturbations, can stabilize these elusive phases; this was recently observed in the three-fold degenerate ground state of a spin-1 atomic gas~\cite{evrard_2021}. However, the resulting state was far from exhibiting complete fragmentation, as quantified by the ``fragmentation entropy'', i.e.~the von Neumann entanglement entropy computed from the eigenvalues of the one-body density matrix. In a $L$-mode system, this entropy is maximized by the $L$-tuple Fock state $\ket{N/L} \equiv \bigotimes_{i=1}^L \ket{N/L}_i$, evenly distributing the bosons in each mode.

A tight-binding lattice model, as realized by cold atoms in optical lattices~\cite{bloch_2008}, naturally offers a set of $L$ orbitals. By activating hopping processes between the sites, the degeneracy of the single-particle spectrum is lifted through the formation of a band structure. As a consequence, bosons condense in the lowest-energy Bloch state, forming a superfluid state over the lattice. Ramping up strong interactions can then fragment the superfluid state, reaching the maximally fragmented Fock state $\ket{N/L}$ in the strongly-correlated limit~\cite{Fabrice_Mott,bloch_2008}. However, this transition from the superfluid to the Mott insulator~\cite{jaksch_1998} has only been realized for limited filling factors $N/L \!\sim\! 1$~\cite{greiner_2002,widera_2005,gerbier_2006,bloch_2008,lepoutre_2019}.

Considering the case $L\!=\!2$, the twin-Fock 
state $\ket{N/2}$ is the most resourceful Fock state for quantum sensing and metrology~\cite{holland_1993, kim_1998, huang_2008, lucke_2011, zou_2018, pezze_2018, barbieri2022}. A multitude of approaches have been proposed and implemented to perform number squeezing around $\ket{N/2}$ (either in true two-mode systems, or in a sub-manifold of spinor gases), such as feedback on non-demolition measurements~\cite{kuzmich_1998, molmer_1999}, parametric amplification~\cite{bucker_2011,lucke_2011}, adiabatic and quasi-adiabatic transformations~\cite{orzel_2001,li_2007,jo_2007,sebby_strabley_2007,rodriguez_2007,rodriguez_2008,esteve_2008,maussang_2010, zhang_2013,luo_2017,evrard_2021, zhang_2024}, shortcuts to adiabaticity~\cite{juliadiaz_2012,yuste_2013}, optimal control~\cite{grond_2009,pichler_2016}, reinforcement learning~\cite{guo_2021} and more~\cite{dunningham_2001,ebert_2014,xin_2023}. Nevertheless, the preparation of a pure twin Fock state in a two-mode system has not yet been achieved.

In this Letter, we explore and compare schemes for fragmenting an ensemble of bosons in a $L$-mode Bose-Hubbard system. We first discuss an adiabatic approach, which consists in slowly ramping up the ratio between Hubbard interactions and hopping amplitude, either directly~\cite{spekkens_1999,greiner_2002} or effectively using a Floquet drive~\cite{eckardt_2005b,lignier_2007}. 
We then introduce a large energy detuning (or ``tilt'') between the modes, which effectively annihilates the hopping.
On the onset of that regime, we use quantum optimal control~\cite{glaser_2015,koch_2022,ansel_2024,lapert_2012} to steer the ground state of the unbiased system towards the $L$-tuple Fock state.
Our results are illustrated for $L\!=\!2$ modes; see~\footnote{\label{suppmat}See Supplementary Material for details and additional results in $L\!=\!2$ and $L\!=\!3$.} for the $L\!=\!3$ case.
With this practical approach, we achieve extreme fragmentation at high filling, two to three orders of magnitude faster than adiabatic methods, and with significant robustness to parameter fluctuations, opening up the prospect of experimental realization.

\begin{figure*}[t]
\begin{center}
\includegraphics[scale=1]{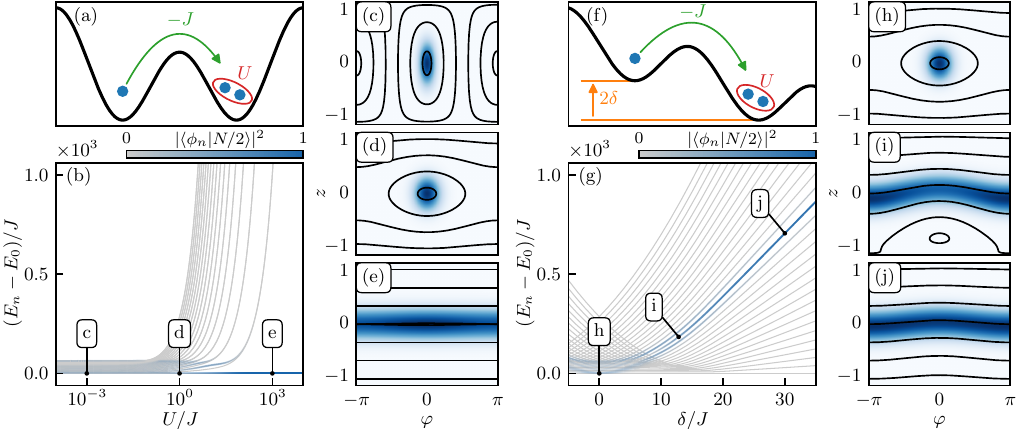}
\caption{
\textbf{Fragmentation in a two-mode Bose-Hubbard system.} 
\textbf{(a)} Schematics of the double-well Hamiltonian~\eqref{eq:Hbjj2}, with hopping $J$, 2-body contact interactions $U$ and no tilt $\delta\!=\!0$. \textbf{(b)} Many-body spectrum for $N\!=\!32$ bosons as a function of $U/J$ (grey lines), with $E_n$ the eigenenergy of eigenstate $\ket{\phi_n}$. The projection of $\ket{N/2}$ onto $\ket{\phi_n}$ is encoded in shades of blue on $E_n$. \textbf{(c-e)} Husimi representations~\cite{husimi_1940,agarwal_1981,haas_2014,pezze_2018} of the states identified in (b) (blue areas, with colorscale ranging from 0 to the maximum value of the distribution) and corresponding MF trajectories (black lines). \textbf{(f-j)} same as (a-e) as a function of the tilt $\delta/J$ with $U/J\!=\!1$.
}
\label{fig1}
\end{center}
\end{figure*}
\paragraph{The system.} We consider a fixed number $N$ of bosons in a two-mode system, such as the double well illustrated in Fig.~\ref{fig1}(a). Under the low-energy single-band approximation, we describe the system by the two-mode Bose-Hubbard model:
\begin{align}
    \label{eq:Hbjj1}
    \hH(t) = &-J\left( \ha^\dagger_1\ha_2 + \ha^\dagger_2\ha_1 \right) + \dfrac{U}{2} \left( \ha^\dagger_1\ha^\dagger_1\ha_1\ha_1 + \ha^\dagger_2\ha^\dagger_2\ha_2\ha_2 \right) \notag\\
    &+ \delta \left( \ha^\dagger_1\ha_1 - \ha^\dagger_2\ha_2 \right),
\end{align}
where $\ha_i^{\dagger}, \ha_i$ are the bosonic creation and annihilation operators in mode $i$, $J$ is the particle hopping strength, $U$ quantifies the onsite two-body interaction and $\delta$ is an eventual tilt between the modes. Using the Schwinger (angular-momentum) representation, 
$\hJ_x\!=\!(\ha_1^\dagger\ha_2 + \ha_2^\dagger \ha_1)/2$, 
$\hJ_y\!=\!(\ha_1^\dagger\ha_2 - \ha_2^\dagger\ha_1)/2\di$, 
$\hJ_z\!=\!(\ha_1^\dagger\ha_1 - \ha_2^\dagger\ha_2)/2$,
Hamiltonian~\eqref{eq:Hbjj1} is rewritten as
\begin{equation}
    \label{eq:Hbjj2}
    \hH(t) = U \hJ^2_z - 2J \hJ_x + 2\delta \hJ_z,
\end{equation}
up to the constant term $U\hN(\hN-2)/4$ (with $\hN = \ha_1^\dagger\ha_1 + \ha_2^\dagger\ha_2$), describing our system as an effective spin-$1/2$ quantum gas on the Bloch sphere. For a sufficiently large $N$, intuition can be built on the quantum dynamics governed by Eq.~\eqref{eq:Hbjj2} by taking the semiclassical, mean-field (MF) limit $\ha_i \rightarrow \sqrt{n_i}\de^{-\di \theta_i}$, yielding the MF Hamiltonian
\begin{equation}
    \label{eq:HMF}
    H(\varphi,z) = \frac{\Lambda}{2}z^2 - \sqrt{1-z^2}\cos{(\varphi)} + \frac{\delta}{J}z,
\end{equation}
where $\Lambda=NU/2J$, and where we have introduced the two canonically conjugate Bloch-sphere variables:~the relative phase $\varphi=\theta_1-\theta_2$ and population $z=(n_1-n_2)/N$~\cite{smerzi_1997,goldman_2023}.

To quantify the fragmentation of an arbitrary state $\ket{\psi}$, we define the following three metrics: (i) The fragmentation (von Neumann) entropy $S_\dF = \sum_i -\lambda_i \log{\lambda_i}$~\cite{raventos2017}, with $\lambda_i$ the normalized eigenvalues of the reduced one-body density matrix with elements $\hat{\rho}_{i,j} \!=\! \bra{\psi} \ha_i^\dagger\ha_j \ket{\psi}$. In a $L$-mode system, $S_\dF$ is maximized to $\ln(L)$ by the $L$-tuple Fock state $\ket{N/L}$. (ii) The number squeezing parameter $\xi_N^2 = \Delta \hJ^2_{z,\psi}/\Delta \hJ^2_{z,\dref}$, quantifying the suppression of particle-number fluctuations with respect to the uncorrelated, binomial distribution of variance 
\mbox{$\Delta \hJ^2_{z,\dref} = N/4$} for $L\!=2\!$~\cite{pezze_2018,Note1}. 
(iii) The quantum fidelity to the twin Fock state $\mathcal{F}=|\braket{N/2}{\psi}|^2$.

\paragraph{Adiabatic approach.} We first consider the preparation of $\ket{N/2}$ by adiabatically increasing $U/J$, as traditionally performed to induce a superfluid to Mott-insulator transition~\cite{jaksch_1998,greiner_2002}. In Fig.~\ref{fig1}(b), we show the many-body spectrum of the Hamiltonian in Eq.~\eqref{eq:Hbjj2} for $N\!=\!32$ and $\delta=0$ as a function of $U/J$. The physics is that of the standard bosonic Josephson junction~\cite{legett_2001,gati_2007}:~Starting from $U/J\ll1$, the ground state $\ket{\phi_0}$ is a superfluid coherent spin state~\cite{agarwal_1981,pezze_2018}, pointing along $(\varphi,z)=(0,0)$ on the Bloch sphere [Fig.~\ref{fig1}(c)]; as one increases interactions, this state asymptotically tends towards $\ket{N/2}$ in the limit $U/J\rightarrow\infty$ [Fig.~\ref{fig1}(d,e)]. Crucially, very large values of $U/J$ are needed to achieve high ground-state fragmentation:~considering $N\!=\!32$ bosons, we find that reaching a fidelity $\mathcal{F}>0.99$ requires $ U/J \gtrsim 230$, while $S_\dF/\ln(2)>0.99$ is reached for $ U/J \gtrsim 285$.

When adiabatically following the states (c) to (e) in Fig.~\ref{fig1}(b), we find that the narrowest energy gap that could trigger Landau-Zener transitions is $\Delta E\!=\!4J$ at $U/J\!=\!0$, after which $\Delta E$ keeps increasing monotonically; this observation does not depend on the number of bosons $N$. An adiabatic preparation of the target state thus requires a preparation time $t_\df \gg \hbar/J$~\cite{Note1}.

As previously noted, total fragmentation requires $U/J\rightarrow\infty$, which, in practice, would correspond to progressively annihilating the hopping strength, e.g.~by increasing the distance between the sites. In an experiment, this could be hard to achieve without exciting motional degrees of freedom~\cite{schumm_2005, grond_2009, maussang_2010, zhang_2024}. Alternatively, $J$ could be effectively suppressed by means of Floquet engineering, using a fast modulation of the tilt $\delta$~\cite{eckardt_2005b,lignier_2007}. We discuss the adiabatic preparation of the $\ket{N/2}$ state using this Floquet approach in~\cite{Note1}; we demonstrate the applicability of this method deep in the high-frequency regime, which is found to be only accessible for limited number of bosons $N$, due to the interacting nature of this periodically-driven system~\cite{Note1}.

\paragraph{Adding a tilt.} To overcome the severe limitations of the adiabatic approach, we now add a strong tilt $\delta$ to the system [Fig.~\ref{fig1}(f)]; for the sake of illustration, we henceforth fix the ratio $U/J$ to a realistic value of 1. In the limit $\delta \gg J,U$, the tilt effectively annihilates the hopping term, as one can deduce by moving to a rotating frame generated by the unitary transformation $\textstyle \hR = \exp \lbrace \di 2\delta \hJ_z t/\hbar\rbrace$; see Refs.~\cite{li_2007,Note1}. In this frame, the eigenstates of the Hamiltonian~\eqref{eq:Hbjj2} become the Fock states $\ket{n_1}\equiv\ket{n_1,N-n_1}$, i.e.~the eigenstates of $\hJ_z$, with the intuitive lab-frame ground state $\ket{0,N}$ due to the large tilt. In Fig.~\ref{fig1}(g), we show the many-body spectrum associated with the Hamiltonian~\eqref{eq:Hbjj2} as a function of $\delta$, for $N\!=\!32$ bosons, identifying the projection of the target twin Fock state $\ket{N/2=16}$. We see that for $\delta/J\!\sim\!1$, the twin Fock state is projected onto many eigenstates of the system, while it indeed becomes an eigenstate for $\delta/J \gg 1$. As $\delta$ increases, the MF fixed-point of lowest energy migrates towards the South pole $z=-1$ [Fig~\ref{fig1}(h-i)]. The remaining trajectories wraps around the Bloch sphere, as do the eigenstates, turning into Fock states of definite relative population $\langle \hJ_z \rangle$. 

From this brief study, summarized in Fig.~\ref{fig1}, one apprehends how difficult it is to adiabatically connect the eigenstates of panels (h) to (j), hence suggesting the development of a more sophisticated approach.

\paragraph{Optimal control.} We now consider optimal control as a promising alternative to prepare twin Fock states $\ket{N/2}$, through an arbitrary variation of the tilt $\delta(t)$ between the wells. To obtain the optimal $\delta(t)$ that steers the ground state $\ket{\phi_0}$ of~\eqref{eq:Hbjj2} (taken at $\delta=0$) towards $\ket{N/2}$ in a control duration $t_\df$ (fixed beforehand, see below), we employ a first-order gradient-based algorithm~\cite{khaneja_2005,werschnik_2007,boscain_2021,dupont_2021,dupont_2023,ansel_2024}. This approach amounts to performing a gradient ascent on a state-preparation metric, which we choose here to be the fidelity $\mathcal{F}$ to the target Fock state $\ket{N/2}$~\cite{Note1}. The quality of the optimal control fields (in terms of e.g. fidelity maximization, robustness to noise and experimental feasibility) yielded by such numerical approaches strongly depends on the initial guess provided as input. For our purposes, we find that a rather large, constant tilt $\delta_0/J \sim \Lambda/2 = NU/4J$ is an efficacious initial guess, as it amounts to a quench into a configuration where both the initial state and the target Fock state are projected into the same restricted subspace consisting of only a few quasi-Fock eigenstates [see the blue states in Fig.~\ref{fig1}(g)], effectively reducing the Hilbert space dimension into which the optimization takes place. In the large $\delta_0$ limit, the spacing between two successive Fock eigenstates $\ket{n_1},\ket{n_1+1}$ 
is $\Delta E \sim U(2n_1-N+1)+2\delta_0\sim 2\delta_0$ for $n_1=N/2$~\cite{Note1}. The period $T_0$ at which a non-steady state evolves in the vicinity of $\ket{N/2}$ in Hilbert space is therefore $T_0 \sim h/2\delta_0$; this establishes a lower  bound on the time associated with state alteration near the equator of the Bloch sphere in the strongly tilted system. Here, we typically opt for a control duration $t_\df\sim~10\,T_0\sim20\,h/NU$~\footnote{Working with finite $\delta_0$, we roughly adjust $t_\df$ and $\delta_0$ around the asymptotic expressions given in the text in order to maximize the fidelity resulting from the initial guess.}. This favorable scaling with $N$ stems from the large tilt $\delta_0 \sim NU/4$ used in our approach. In a tight-binding framework, relevant to optical lattices~\cite{eckardt_2017}, this would ultimately limit the maximum number of bosons $N$ that one can consider.

\begin{figure}[t]
\begin{center}
\includegraphics[scale=1]{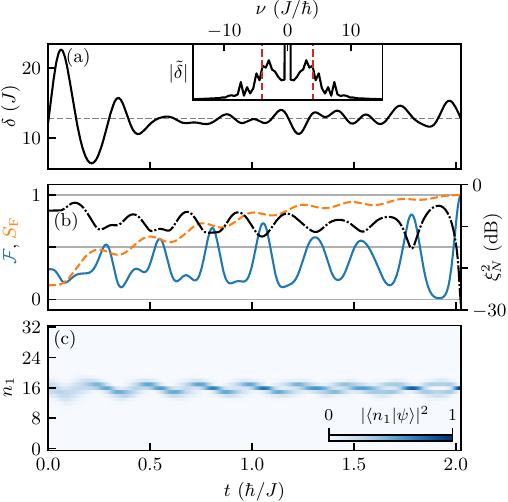}
\caption{
\textbf{Twin Fock state preparation by optimal control.} \textbf{(a)}~Optimal control tilt as a function of time (solid black line) to drive the ground state of Eq.~\eqref{eq:Hbjj2} with $U/J=1$ and $N=32$ (taken at $\delta/J=0$) towards the twin Fock state $\ket{N/2=16}$, and initial guess $\delta(t)=\delta_0$ (dashed grey line). Inset: norm of the Fourier transform of $\delta(t)$ (solid black line, $y$-axis going from 0 to $1J$) and $\pm1/T_0$ (dashed red lines, see text). \textbf{(b)}~Fidelity $\mathcal{F}$ to $\ket{N/2=16}$ (blue solid line), fragmentation entropy $S_\dF$ (divided by $\ln(2)$, orange dashed) and number squeezing parameter $\xi^2_N$ (dB scale, black dash-dotted) as a function of time. See Table~\ref{tab1} for details. \textbf{(c)}~Fock space projection during the preparation.}
\label{fig3}
\end{center}
\end{figure}

Numerical results for the preparation of the twin Fock state, using optimal control of the tilt, are shown in Fig.~\ref{fig3} for $N=32$ and $U/J=1$. Panel (a) shows the optimal $\delta(t)$, displaying a fast, non-trivial oscillation around $\delta_0$. This is a general trend observed for different $N$ values, and a Fourier analysis of $\delta(t)$ [inset of Fig.~\ref{fig3}(a)] reveals that its main non-zero frequencies are grouped around the transition frequency $\Delta E/h = 1/T_0$ between two adjacent equatorial Fock states in the large $\delta_0$ limit. The algorithm thus naturally converges on a quasi-periodic modulation that hybridizes neighboring quasi-Fock eigenstates, in order to then maximize the projection onto the targeted Fock state. The resulting fidelity $\mathcal{F}$, fragmentation entropy $S_\dF$ and number squeezing $\xi_N^2$ over the preparation are plotted on Fig.~\ref{fig3}(b), while (c) features the Fock-space projection of the prepared state. For those parameters, our metrics finally reach $\mathcal{F} = 0.999$ (breaking condition of the iterative algorithm), $S_\dF/\ln(2) = 0.9998$ and $\xi_N^2 = 1.33\cdot10^{-3}=-28.8$~dB, i.e. number-difference fluctuations suppressed by nearly 3 orders of magnitude compared with coherent spin states lying on the equator of the Bloch sphere~\cite{pezze_2018}. One observes how, for this strongly non-adiabatic scheme, the fidelity to the most fragmented state $\ket{N/2}$ is a much more fragile figure of merit for fragmentation than the fragmentation entropy. This can be appreciated by considering, for instance, that the very next Fock states $\ket{N/2\pm1}$ have high fragmentation entropy but zero fidelity to $\ket{N/2}$. Results from $N=4$ to 512 are compiled in Table~\ref{tab1}. Apart from the fidelity, which decreases appreciably as $N$ increases, we see that optimal control allows to reach high fragmentation entropy and number squeezing up to substantially large $N$. It should be noted that, as $U/J$ is fixed in this study, the ground state of the unbiased Hamiltonian~\eqref{eq:Hbjj2} gets inherently number squeezed as $N$ increases~\footnote{As the mean-field interaction parameter $\Lambda\propto NU/J$, increasing $N$ at fixed $U/J$ gets us deeper into the Josephson regime~\cite{legett_2001,gati_2007,pezze_2018}} (see Table~\ref{tab1}; the number-squeezing gain due to optimal control being the difference between the last two rows). Overall, we see that optimal control combined with a strong average tilt $\delta_0$ is a promising method to produce the maximally fragmented Fock states $\ket{N/2}$ in a two-mode Bose-Hubbard system;~see~\cite{Note1} for similar results on the $L\!=\!3$ case. Moreover, one achieves drastic number squeezing, two to three orders of magnitudes faster than (quasi-)adiabatic methods; see Refs.~\cite{orzel_2001,li_2007,jo_2007,sebby_strabley_2007,esteve_2008,zhang_2013,luo_2017} and~\cite{Note1}.
\newcommand\cwa{0.16}
\newcommand\cwb{0.0855}
\begin{table}[t]
\begin{center}
\begin{tabular}{
|>{\centering}p{\cwa\linewidth}|
>{\centering}p{\cwb\linewidth}
>{\centering}p{\cwb\linewidth}
>{\centering}p{\cwb\linewidth}
>{\centering}p{\cwb\linewidth}
>{\centering}p{\cwb\linewidth}
>{\centering}p{\cwb\linewidth}
>{\centering}p{\cwb\linewidth}
>{\centering\arraybackslash}p{\cwb\linewidth}|
}
\hhline{=========}
$N$ & 4 & 8 & 16 & 32 & 64 & 128 & 256 & 512 \\ \hline
$\delta_0 (J)$ & 2.80 & 4.40 & 7.20 & 12.8 & 24.0 & 41.6 & 64.0 & 89.6 \\
$t_\df\,(\hbar/J)$ & 5.05 & 3.43 & 2.77 & 2.02 & 1.47 & 0.925 & 0.702 & 0.500 \\
$\mathcal{F}$ & \multicolumn{4}{c}{\; \dhrulefill \; $0.999$ \; \dhrulefill \;} & 0.987 & 0.951 & 0.923 & 0.888 \\
$S_\dF/\ln(2)$ & \multicolumn{4}{c}{\; \dhrulefill \; $>0.999$ \; \dhrulefill \;} & 0.994 & 0.980 & 0.959 & 0.950 \\
\multirow{2}{*}{
\hspace{-2mm}$\begin{array}{c} \xi^2_N\\ \text{(dB)}\end{array}\hspace{-2mm}\left(\hspace{-1mm}\begin{array}{l} \psi_\df \\ \phi_0\end{array}\hspace{-1mm}\right)$
} & -26.0 & -22.6 & -21.6 & -28.8 & -19.0 & -14.7 & -15.2 & -19.3 \\
& -2.30 & -3.52 & -4.84 & -6.23 & -7.66 & -9.12 & -10.6 & -12.1 \\
\hhline{=========}
\end{tabular}
\caption{
\textbf{Twin Fock state preparation by optimal control} ($U/J=1$) for increasing number of bosons $N$, with guess tilt $\delta_0$, control duration $t_\df$, final fidelity $\mathcal{F}$ to $\ket{N/2}$, fragmentation entropy $S_\dF$ and number squeezing parameter of prepared and initial states ($\ket{\psi_\df}$ and $\ket{\phi_0}$ resp.).
}
\label{tab1}
\end{center}
\end{table}

The full quantum dynamics depends on the details of the instantaneous many-body spectra, over the entire control duration. A priori, a limitation of this approach, as compared to adiabatic schemes, is that it requires a precise knowledge of the number of bosons $N$. Quite generically, optimal control lacks robustness against parameter fluctuations, especially those associated with the particle number $N$~\cite{lapert_2012}. However, we find that in this large-average-tilt configuration (where the eigenenergy spacing between successive quasi-Fock eigenstates loses its dependence on the average number of particles in said eigenstates), an optimal $\delta(t)$ computed for a given $N$ remains efficient to fragment the Bose gas for a close but different $N'=N+2k$ (keeping commensurability with $k\in\mathbb{Z}^*$), and can be used as a solid initial guess for another optimization with the correct $N'$~\cite{Note1}.

\begin{figure}[!t]
\begin{center}
\includegraphics[scale=1]{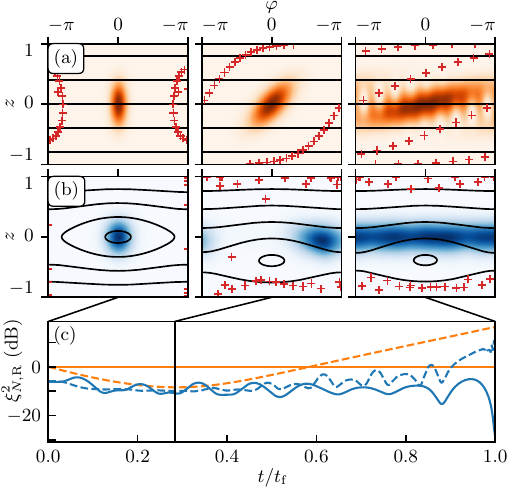}
\caption{
\textbf{Optimized axis twisting}  ($N=32$). \textbf{(a,b)} Husimi state representations (colored area), their zeros (Majorana anti-stars, red crosses) and instantaneous MF trajectories (solid black lines) on the Bloch sphere. (a) correspond to standard one-axis twisting of a coherent spin state, with initial condition (left), maximum spin-squeezing time (middle) and oversqueezing at a longer time (right). (b) is the optimal preparation of $\ket{N/2}$ featured in Fig.~\ref{fig3} for three different increasing times (left to right). \textbf{(c)} Number squeezing $\xi^2_N$ (solid lines) and spin squeezing $\xi^2_\dR$ (dashed lines) of (a,b) (resp. orange, blue) over the protocols (of independent duration $t_\df=2N^{-1/2}\hbar/U$ (a) and $t_\df=2.02\,\hbar/J$ (b)).
}
\label{fig4}
\end{center}
\end{figure}

In Fig.~\ref{fig4}, we further interpret the algorithm's ``strategy'' in performing number squeezing around $\ket{N/2}$ in this biased two-mode system by drawing a parallel with spin-squeezing performed by one-axis twisting (OAT)~\cite{kitagawa_1993,micheli_2003,pezze_2018}. Through OAT, a coherent spin state initially pointing along $\vec{n}$ on the Bloch sphere's equator undergoes \textit{spin-squeezing} through an evolution governed by the interaction Hamiltonian $U \hJ^2_z$, due to MF trajectories flowing in opposite directions on each Bloch hemisphere ($\dd\varphi/\dd t\propto z$). The resulting spin squeezing, occurring along a tangential direction $\vec{s}$ to the average spin direction $\vec{n}$, can be quantified by the squeezing parameter $\xi_\dR^2 = N (\Delta \hJ_{\vec{s}})^2 / \langle \hJ_{\vec{n}} \rangle^2$~\cite{wineland_1994,pezze_2018}, assessing the metrological usefulness of the anisotropy of quantum fluctuations around $\vec{n}$. OAT spin-squeezes an equatorial coherent spin state for $t \lesssim N^{-1/2}\hbar/U$, while for longer times, it gets over-squeezed as it wraps around the Bloch sphere, resulting in $\xi_\dR^2>1$~\cite{kitagawa_1993,pezze_2018} [Fig.~\ref{fig4}(a,c)]. A salient symptom of over-squeezing is the constellation of zeros riddling the Husimi function [Fig.~\ref{fig4}(a)], which are Majorana anti-stars, antipodes of the $N$ individual $1/2$-spins into which the state can be factored according to the Majorana stellar representation~\cite{majorana_1932,bloch_1945,bruno_2012,evrard_2019}. However, OAT never number squeezes, having Fock states as its eigenstates [Fig.~\ref{fig4}(c)]. In the case of optimal control [Fig.~\ref{fig4}(b)], the large tilt yields MF trajectories similar to OAT near the equator, resulting in a slight short-term spin squeezing (panel (c)). The control  on $\delta(t)$ then constrains the state towards the equator while it wraps around the Bloch sphere, gradually sending the Majorana anti-stars to its poles. Our optimal control approach, operating around a large tilt value, can thus be viewed as number squeezing through optimized axis twisting.

\paragraph{Conclusion.} In this work, we introduced a practical and efficient approach to achieve total fragmentation of a Bose gas in the $L$-mode Bose-Hubbard model, i.e. to prepare the evenly distributed $L$-tuple Fock state $\ket{N/L,...,N/L}$ with $N$ bosons. We illustrated our results for systems with $L\!=\!2$ and 3 modes. We first considered the conventional adiabatic ramping of the ratio between interaction $U$ and hopping $J$ (either directly varied, or effectively via Floquet engineering), finding strong limitations and constraints:~long preparation time, low particle number and high Floquet frequency. We then applied quantum optimal control with the energy bias $\delta$ between the modes as our control parameter. We found that by operating around a large tilt $\delta \approx NU/4$, one is able to diabatically prepare the twin Fock state in a duration $t_\df \approx 20\,\hbar/NU$, with appreciable robustness against atom number fluctuations~\cite{Note1}. This work offers realistic approaches to experimentally achieve $L$-tuple Fock state preparation, the most resourceful number state for metrological purposes~\cite{pezze_2018}.
We emphasize that our scheme relies on modulating the tilt (or detuning between the modes) around a large mean value. This can be advantageous for certain physical settings that cannot operate close to resonance (e.g.~schemes using two coupled internal states operating in the large-detuning regime).

\paragraph{Acknowledgements.} The authors thank Jean Dalibard for insightful discussions, and for his careful reading of the manuscript. We also acknowledge discussions with Markus Oberthaler, Helmut Strobel, Maximilian Pr\"ufer, Leon Zaporski and Shai Tsesses, as well as support from the ERC (LATIS project) and the EOS (CHEQS project).

\bibliography{bibliography}

\clearpage
\onecolumngrid
\setcounter{equation}{0}
\setcounter{figure}{0}
\setcounter{table}{0}
\renewcommand{\theequation}{S\arabic{equation}}
\renewcommand{\thefigure}{S\arabic{figure}}
\renewcommand{\thetable}{S\arabic{table}}
\renewcommand{\thesection}{S\arabic{section}}
\renewcommand{\thesubsection}{\thesection.\arabic{subsection}}

\begin{center}
	{\Large\bfseries Supplementary Material}
\end{center}

\section{A.~ Fragmentation from adiabatic ground-state following}
\label{app:direct_UJ_ramping}
We consider the fragmentation of the many-body ground state in the $L$-mode Bose-Hubbard Hamiltonian
\begin{equation}
    \label{eq:Hbjj_UJ}
    \hH = -J \sum_{\ell=1}^{L-1} \left(\ha^\dagger_\ell\ha_{\ell+1} + \ha^\dagger_{\ell+1}\ha_{\ell}\right) + \dfrac{U}{2}\sum_{\ell=1}^{L} \ha^\dagger_\ell\ha^\dagger_\ell\ha_\ell\ha_\ell,
\end{equation}
as one tunes the ratio $U/J$; see Fig.~1 of the main text. In this work, we focus on the two-mode ($L=2$) and three-mode ($L=3$) cases.

\begin{figure}[!b]
\begin{center}
\includegraphics[scale=1]{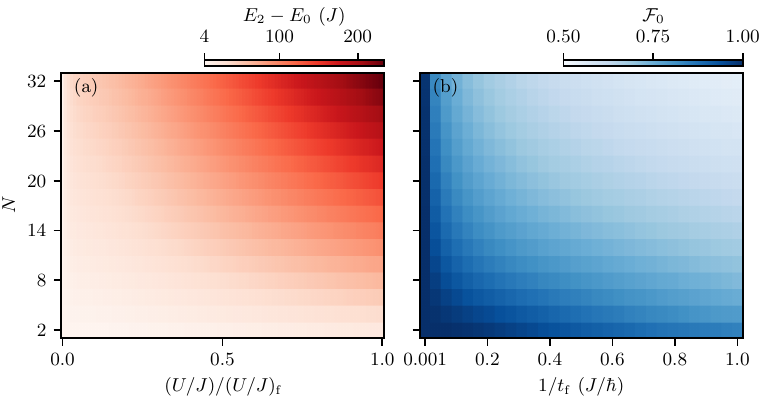}
\caption{
\textbf{Adiabatic ground state following as $U/J$ increases.} 
\textbf{(a)}~Energy gap between the ground state and the first accessible excited state as a function of $U/J\in[0,(U/J)_\df]$ (see text) and $N\!\in\![2,32]$ (only considering even integers for commensurability). 
\textbf{(b)}~Final fidelity $\mathcal{F}_0=|\langle \phi^{(U/J)_\df}_0 | \psi(t_\df) \rangle|^2$ (see text) after a linear ramping of $U/J$ from 0 to $(U/J)_\df$ and as a function of the ramping rate $1/t_\df$ and $N$. The initial state is $|\phi^{(0)}_0 \rangle$.
}
\label{fig:adiab_UJ}
\end{center}
\end{figure}

First for $L\!=\!2$, one can rewrite the Hamiltonian $\hH = U\hJ^2_z - 2J\hJ_x$ with $\hJ_x = (\ha_1^\dagger\ha_2+\ha_2^\dagger\ha_1)/2$ and $\hJ_z = (\ha_1^\dagger\ha_1-\ha_2^\dagger\ha_2)/2$. Considering $N$ bosons, we denote $E^{(U/J)}_n$ and $|\phi^{(U/J)}_n\rangle$ the $N+1$ eigenenergies and eigenstates of~\eqref{eq:Hbjj_UJ} as a function of $U/J$. Starting from $U/J\ll1$, the hopping term dominates, and we have
\begin{align}
-2J\hJ_x &= -J(\ha^\dagger_1\ha_2 + \ha^\dagger_2\ha_1)\notag\\
&= -J(\ha^\dagger_+\ha_+ - \ha^\dagger_-\ha_-),
\end{align}
with $\ha_\pm = (\ha_1 \pm \ha_2)/\sqrt{2}$. In the limit $U=0$, the eigenstates are thus of the form
\begin{equation}
    \label{eq:gs_jx}
    \ket{\phi^{(0)}_n} = \dfrac{\left(\ha_+^\dagger\right)^{N-n}}{\sqrt{(N-n)!}}\dfrac{\left(\ha_-^\dagger\right)^{n}}{\sqrt{(n)!}} \ket{\varnothing},
\end{equation}
with $E_n^{(0)}=J(2n-N)$ and where $\ket{\varnothing}$ is the vacuum state. The total Hamiltonian (hopping terms and interactions) is invariant under the swapping of modes $1\leftrightarrow2$, which is associated with the parity operator $\hP$ acting as $\hP\ha_\pm = \pm \ha_\pm$. From Eq.~\eqref{eq:gs_jx}, we see that the eigenstates of $\hJ_x$ are directly sorted in subspaces of opposite parity, through the parity of $n$: $\hP|\phi^{(U/J)}_n\rangle=(-1)^n|\phi^{(U/J)}_n\rangle$. Starting from $|\phi^{(0)}_0\rangle$ and given that $\hH$ does not couple the two subspaces of opposite parity $\forall\,U/J$, the first possible excitation at $U/J=0$ is towards $|\phi^{(0)}_2\rangle$, with energy $\Delta E^{(0)} = E^{(0)}_2-E^{(0)}_0 = 4J$ (Fig.~\ref{fig:adiab_UJ}(a)). This gap enlarges as $U/J$ increases, up to $\Delta E \sim U$ between $\ket{\phi_0}=\ket{N/2}$ and $\ket{\phi_2}=(\ket{N/2+1}+\ket{N/2-1})/\sqrt{2}$ in the strongly interacting regime $U/J\gg1$ (with the notation $\ket{n_1} \equiv \ket{n_1}_1\otimes\ket{N-n_1}_2$).

\begin{figure}[!t]
\begin{center}
\includegraphics[scale=1]{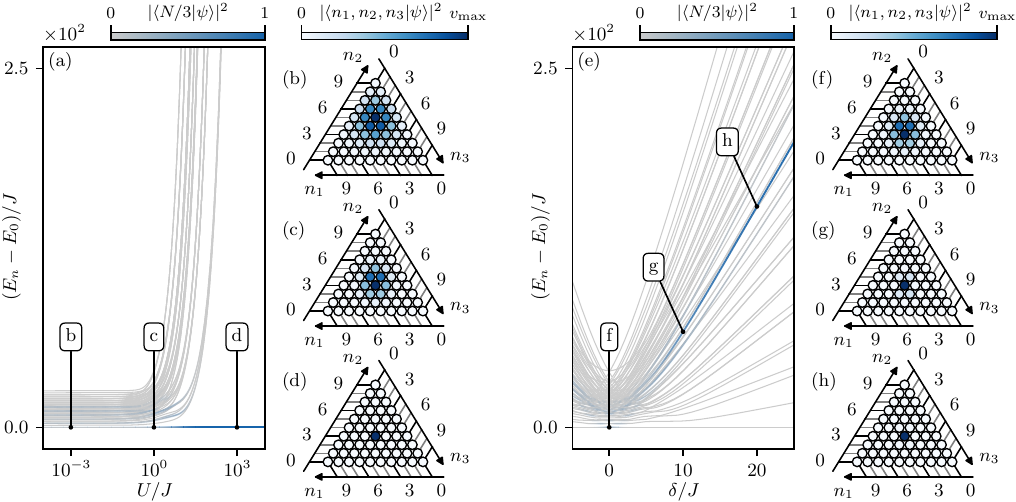}
\caption{
\textbf{Fragmentation in a three-mode Bose-Hubbard system.} 
\textbf{(a)} Many-body spectrum for $N\!=\!9$ bosons as a function of $U/J$ (grey lines), with $E_n$ the eigenenergy of eigenstate $\ket{\phi_n}$. The projection of $\ket{N/3}$ onto $\ket{\phi_n}$ is encoded in shades of blue on $E_n$.
\textbf{(b-d)} Fock state projection of the states identified in (a) (blue areas, with independent maximum values $v_\dmax = 0.0514$ (b), 0.177 (c) and 0.999 (d)).
\textbf{(e-h)} same as (a-d) as a function of the tilt $\delta/J$ with $U/J\!=\!1$, and $v_\dmax = 0.177$ (b), 0.592 (c) and 0.884 (d).
}
\label{fig:3well_principle}
\end{center}
\end{figure}

The maximally fragmented state $\ket{N/2}$ can be reached from $|\phi^{(0)}_0\rangle$ through adiabatic ground-state following by slowly ramping-up $U/J$. The adiabaticity condition on the ramping time can be estimated in first approximation by the inverse of the frequency associated with the narrowest gap in the spectrum, i.e. $t_\df \gg h/4J$. To go beyond this estimate, we perform simulation of adiabatic ground state following as a function of the ramping time $t_\df$ and $N$. Starting from $|\phi^{(0)}_0\rangle$, we linearly ramp up $U/J$ until a value $(U/J)_\df$ (increasing with $N$) such that $|\langle N/2|\phi_0^{(U/J)_\df}\rangle|^2 \approx 0.99$. Figure~\ref{fig:adiab_UJ}(b) shows the ground state fidelity $\mathcal{F}_0=|\langle \phi^{(U/J)_\df}_0 | \psi(t_\df) \rangle|^2$ of the final state as a function of $N$ and $t_\df$. 
While the narrowest relevant gap in the spectrum is $4\,J$ $\forall\,N$, we see that the minimum $t_\df$ required to reach a satisfactory fidelity $\mathcal{F}_0$ increases with $N$. This is due to the increasing number of levels in the spectrum that provide additional channels towards which the evolved state can propagate. We see that the lowest ramping time to get $\mathcal{F}_0\approx1$ is already of the order or $10^3 \hbar/J$ for $N\!=32$.

Regarding the $L\!=\!3$ case: we display in Fig.~\ref{fig:3well_principle} a plot analogous to Fig.~1 (main text for $L=2$). Similarly to the case $L\!=\!2$, adiabatically increasing $U/J\rightarrow\infty$ allows one to reach the maximally fragmented state $\ket{N/3}\equiv\ket{N/3,N/3,N/3}$ from the ground state of the system with finite interactions (panels (a-d). There, the scaling of the Hilbert space dimension $d_\dH$ with $N$ is quadratic: $d_\dH = (N+1)(N+2)/2$, which, as discussed previously, is expected to hinder adiabaticity, even more than for $L\!=\!2$. In panels (e-h), we show that, as for $L\!=\!2$, an additional constant tilt reveals the target state in the spectrum. In Sec.~\hyperref[app:qoc]{D}, we utilize this tilt as a parameter to prepare $\ket{N/3}$ by optimal control.

\section{B.~ Floquet control of $U/J$} 
\label{app:floquet}
One can introduce a fast periodic modulation of the tilt around $\delta=0$ to dynamically control the effective tunneling between the two modes~\cite{eckardt_2005b}. In this way, one can drive a superfluid to Mott insulator transition \cite{lignier_2007} by \textit{slowly} ramping down the effective tunneling of the particles.

More precisely, we consider a strong sinusoidal modulation of the tilt $\delta(t)=A\cos(\omega t)$ and the Hamiltonian of the periodically driven system is written as
\begin{align}
    \label{eq:Hbjj_periodic drive}
    \hat{H}(t) = &-J \sum_{\ell=1}^{L-1} \left(\ha^\dagger_{\ell}\ha_{\ell+1} + \ha^\dagger_{\ell+1}\ha_{\ell} \right) + \frac{U}{2} \sum_{\ell=1}^{L} \ha^\dagger_\ell \ha^\dagger_\ell\ha_\ell\ha_\ell + A \cos{(\omega t)} \sum_{\ell=1}^{L} \ell \, \ha^\dagger_\ell\ha_\ell,
\end{align}
Changing to a frame of reference rotating with the same frequency $\omega$, generated by the unitary transformation
\begin{align}
    \hat{R} &= \exp \left\lbrace \di\frac{A}{\hbar} \int_0^{t} \mathrm{d}t' \, \cos{(\omega t')} \sum_{\ell=1}^{L} \ell \, \ha^\dagger_\ell\ha_\ell\right\rbrace,
\end{align}
we obtain a transformed Hamiltonian
\begin{align}
    \tilde{H}(t) &= -J \left( \exp \left\lbrace \di \frac{A}{\hbar\omega} \sin{(\omega t)}\right\rbrace  \sum_{\ell=1}^{L-1} \ha^\dagger_{\ell}\ha_{\ell+1}  + \mathrm{h.c.}\right) + \frac{U}{2} \sum_{\ell=1}^{L} \ha^\dagger_\ell \ha^\dagger_\ell\ha_\ell\ha_\ell.
\end{align}
In the high-frequency approximation, the dynamics of the system can be captured at stroboscopic times --- integer multiples of the modulation  period $T$ --- by a time-independent effective Hamiltonian $\hH_{\mathrm{eff}}$, which can be perturbatively approximated using an infinite series in powers of $\omega^{-1}$ based on the Magnus expansion ~\cite{rahav_2003, goldman_2014,eckardt_2017}. At zeroth order (in the energy frame moving with the periodic tilt \cite{Note1}), the Hamiltonian sees its hopping strength renormalized as ~\cite{eckardt_2005b,lignier_2007}:
\begin{align}
    \label{eq:zeroth_order_eff_H}
    \hH_\deff^{(0)} = \frac{1}{T} \int_{0}^{T} \dd t \, \tilde{H}(t) = \frac{U}{2} \sum_{\ell=1}^{L} \ha^\dagger_\ell \ha^\dagger_\ell\ha_\ell\ha_\ell - J \mathcal{J}_0\left(\frac{A}{\hbar\omega} \right) \sum_{\ell=1}^{L-1} \left(\ha^\dagger_{\ell}\ha_{\ell+1} + \ha^\dagger_{\ell+1}\ha_{\ell} \right). \\
\end{align}
The first order correction is given by
\begin{align}
    \label{eq:first_order_eff_H}
    \hH_\deff^{(1)} &= \frac{1}{2T\di} \int_{0}^{T} \dd t_1 \int_{0}^{t_1} \dd t_2 \, [\tilde{H}(t_1),\tilde{H}(t_2)] \nonumber \\
    &= \frac{UJ\pi}{2\hbar\omega} \mathcal{H}_0\left(\frac{A}{\hbar\omega} \right)  \sum_{\ell=1}^{L-1} \left(\ha^\dagger_\ell \hat{n}_\ell \ha_{\ell+1} - \ha^\dagger_\ell \hat{n}_{\ell+1} \ha_{\ell+1} + \ha^\dagger_{\ell+1} \hat{n}_\ell \ha_{\ell} - \ha^\dagger_{\ell+1} \hat{n}_{\ell+1} \ha_{\ell} \right) \nonumber\\
    &\quad + \frac{J^2\pi}{\hbar\omega} \mathcal{H}_0\left(\frac{A}{\hbar\omega} \right) \mathcal{J}_0\left(\frac{A}{\hbar\omega} \right) \left( \ha^{\dagger}_1\ha_1 - \ha^{\dagger}_L\ha_L\right),
\end{align}
where $\mathcal{J}_0(x)$ is the zeroth order Bessel function and $\mathcal{H}_0(x)$ is the zeroth order Struve function.
In the limit $\hbar\omega \gg J,U$, one can negelct the effect of the first (and higher) order term and approximate $\hH_\deff \approx \hH_\deff^{(0)}$. Under such a scenario, one can engineer a Floquet quasi-adiabatic~\cite{eckardt_2015,novicenko_2017} state preparation scheme~\cite{eckardt_2005b}: starting from the ground state of  $\hH(t)=\hH^{(0)}_\deff$ at $A=0$, one \textit{slowly} ramps up $A/\hbar \omega$ (at fixed $\omega$) up to the first zero $x_0\approx 2.4$ of $\mathcal{J}_0(x)$. Following the ground state of $\hH^{(0)}_\deff$, the resulting state is the maximally fragmented state $\ket{N/L}$, ground state of the purely interacting Hamiltonian.

\begin{figure}[!t]
\begin{center}
\includegraphics[scale=1]{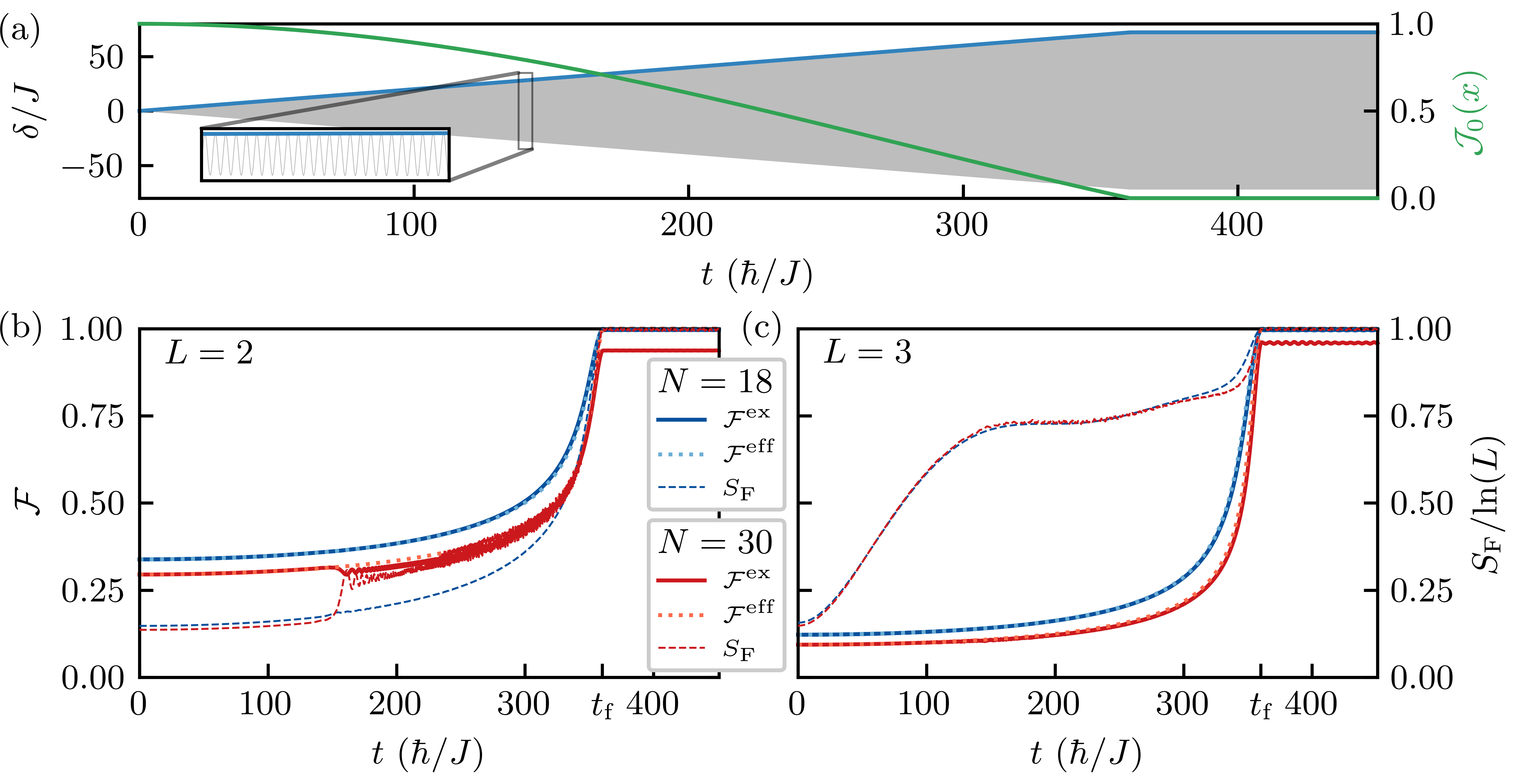}
\caption{
\textbf{Fragmented state preparation by fast modulation of the tilt}, starting from the ground state of~\eqref{eq:Hbjj_periodic drive} with $A=0$ and $U/J=1$. \textbf{(a)} Tilt $\delta(t)=A(t)\cos{(\omega t)}$ (grey line, $\omega=30J/\hbar$) with $A(t)=mt$ (blue envelope, $m=0.2J^2/\hbar$) until $t_\df \approx 2.4 \, \hbar\omega/m$ and constant thereafter. $\mathcal{J}_0(x)$ (green line) with $x=A/\hbar\omega$. \textbf{(b)} Fidelities to the twin Fock state of $\ket{\psi(t)}$ ($\mathcal{F}^\deff$, solid line) and of the ground state of~\eqref{eq:zeroth_order_eff_H} (dotted line), and fragmentation entropy $S_{\mathrm{F}}$ of $\ket{\psi (t)}$ for $N=16$ (blue) and 32 (red).
}
\label{fig:floquet}
\end{center}
\end{figure}

Numerical results for $U/J=1$ and $L = 2$ and $3$ are shown in Fig.~\ref{fig:floquet}. The protocol is drawn in panel~(a), showing $\delta(t)$ with frequency $\omega=30\,J/\hbar$, and envelope $A(t)=mt$ with slope $m = 0.2 J^2/\hbar$, reaching $A(t)=x_0$ at time $t_\df \approx 2.4 \, \hbar\omega/m = 360 \,\hbar/J$. Upon reaching its target value, the amplitude is maintained to observe the realization. The second row features results of numerical simulations for $N=18$ and $30$ at $L=2$ (Panel~(b)) and $L=3$ (Panel~(c)). We show, as a function of time, the fragmentation entropy $S_{\mathrm{F}}$ of the state propagated with the exact Hamiltonian~\eqref{eq:Hbjj_periodic drive}, as well as two fidelities to $\ket{N/L}$: the fidelity $\mathcal{F}^\dex$ of the state $\ket{\psi(t)}$ propagated according to the exact Hamiltonian, and the fidelity $\mathcal{F}^\deff$ of the ground state of $\hH_\deff^{(0)}$~\eqref{eq:zeroth_order_eff_H} (expected to reach 1 $\forall N$). For $N=18$, we achieve exact preparation of $\ket{N/2=9}$, maximizing the final values of $\mathcal{F}^\dex$ and $S_\dF$. For $N=30$ however, while the fragmentation entropy reaches almost $\ln(2)$, the final exact fidelity is not maximized which indicates a breakdown of the effective Hamiltonian approximation.

\begin{figure}[b]
    \centering
    \includegraphics{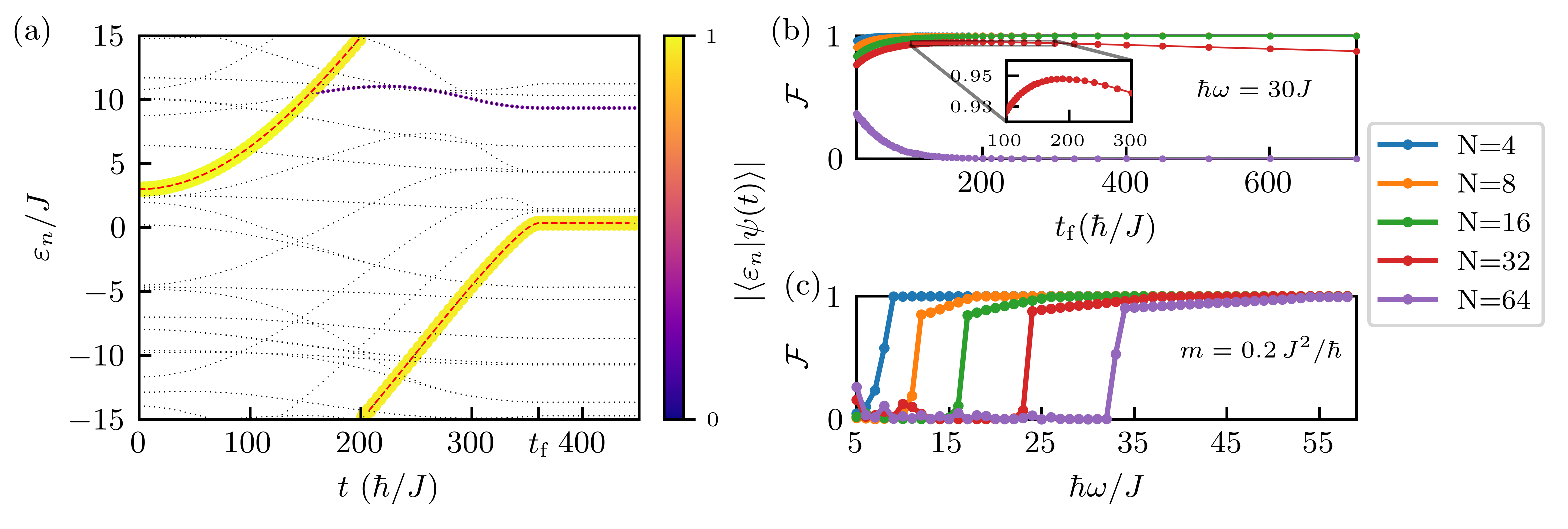}
    \caption{\textbf{(a)} Instantaneous Floquet quasi-energy spectrum $\varepsilon_n$ as a function of the ramping time $t$ of the modulation amplitude (black dots) and quasi-energy of the Floquet state maximizing the overlap with $\ket{N/2}$ (red dashed line) for $N=32$. The projection of $\ket{\psi(t)}$ on the instantaneous Floquet states is encoded in the radius and colorbar shade. \textbf{(b)} Final exact fidelity $\mathcal{F}^\dex$ as a function of the ramp duration $t_\df$ for different $N$ at $\hbar\omega=30J$. \textbf{(c)} $\mathcal{F}^\dex$ as a function of $\omega$ for a fixed ramp slope $m=0.2 J^2/\hbar$}
    \label{fig:Floquet_leaking}
\end{figure}

To understand the reason behind the failure of the effective model, we plot the instantaneous Floquet quasi-energy spectra~\cite{eckardt_2017} of \eqref{eq:Hbjj_periodic drive} as a function of $t$ (and thus $A$) for $L = 2$ and $N = 30$ in panel~(a) of Fig.~\ref{fig:Floquet_leaking} . Following how the prepared state projects in the spectrum, we see that the failure of the effective Hamiltonian to model the exact dynamics corresponds to a leakage towards other, untargeted Floquet states [see purple eigenenergies in Panel (a)]. Indeed, many narrow avoided crossings are to be seen, resulting from the Floquet folding of the static spectrum (a realistic necessity considering the width of the many body spectrum~\cite{eckardt_2017,novicenko_2017}). Hence, as we consider more bosons, the Hilbert space dimension increases (linearly for $L=2$), and with it the number of fortuitous avoided crossings that the prepared state has to cross diabatically~\cite{eckardt_2017}. To achieve better fidelity in many-body ``adiabatic'' Floquet engineering~\cite{eckardt_2015,novicenko_2017}, one must embrace diabaticity and moderately increase the slope of parameter variation to favor Landau-Zener transitions through the inevitable narrow crossings. One can see in Fig.~\ref{fig:floquet}(b) that there exists an optimum fidelity achievable with this quasi-adiabatic Floquet driving, however quite limited in $N$ at fixed $\omega$.

We finally study the requirement on $\omega$ as a function of $N$ (at fixed slope $m$ and $L=2$). In Fig.~\ref{fig:floquet}(c), one can see that the driving frequency $\omega$ required to maintain a high fidelity (i.e. proficiency of the high frequency approximation) scales almost linearly with $N$. This scaling puts a limit on the maximum number of atoms that one can use to create a fragmented state in a realistic cold-atom setup using this method. Indeed, for experimentally realistic parameters, the single-band Bose-Hubbard model is valid for typical energy scales $E\ll100J$~\cite{eckardt_2005a,eckardt_2017}; since the maximum $\omega$ is limited to around $30J$ before the driving strength $A(t_\df) = 2.4 \hbar \omega$ (zero of the Bessel function) reaches $100J$, limiting $N$ to approximately 32 atoms.

\section{C.~ Negating the hopping by a large constant tilt}
\label{app:constant_tilt}
We show how a strong tilt between the modes of the Bose-Hubbard model effectively annihilates the hopping, detailing the derivation for $L=2$ and $L=3$ modes.

\subsection{a.~ $L=2$}
We consider the Hamiltonian 
\begin{equation}
    \label{eq:Hbjj_2well_tilt}
    \hH = U \hJ^2_z - 2J \hJ_x + 2\delta \hJ_z,   
\end{equation}
in the rotating frame accessed by the unitary transformation $|\widetilde{\psi}(t)\rangle = \hR |\psi(t)\rangle$ with $\hR = \text{exp}\lbrace\di 2\delta \hJ_z t/\hbar\rbrace$. In this frame, the Hamiltonian reads
\begin{align}
    \htH(t) &= \hR \hH \hR^\dagger - \di\hbar \hR \dfrac{\dd \hR^\dagger}{\dd t}\notag\\
    &= U\hJ^2_z - 2J \, \de^{\di 2\delta \hJ_z t/\hbar} \hJ_x \, \de^{-\di 2\delta \hJ_z t/\hbar}\notag\\
    &=U\hJ^2_z - 2J\cos\left(\dfrac{2\delta t}{\hbar}\right) \hJ_x + 2J\sin\left(\dfrac{2\delta t}{\hbar}\right) \hJ_y,
\end{align}
using the algebra $[\hJ_i,\hJ_j] = \di\epsilon_{ijk}\hJ_k$. If $\delta \gg U,J$, the Hamiltonian $\htH$ can be approximated by its Magnus expansion at zeroth order, namely its average over one period $\textstyle \htH \approx \htH_\deff^{(0)} = U\hJ^2_z$. Finally, going back to the initial frame gives
\begin{align}
    \hH &= \hR^\dagger \htH \hR - \di \hbar \hR^\dagger\dfrac{\dd \hR}{\dd t} \notag\\
    &\approx U\hJ^2_z + 2 \delta \hJ_z,
\end{align}
whose eigenstates are the Fock states, among which the maximally fragmented state $\ket{N/2}$; see Fig.~1 of the main text.

\subsection{b.~ $L=3$}
We consider the titled Bose-Hubbard Hamiltonian,
\begin{align}
    \label{eq:Hbjj_3well_tilt}
    \hat{H}(t) &= -J \sum_{\ell=1}^{2} \left(\ha^\dagger_{\ell}\ha_{\ell+1} + \ha^\dagger_{\ell+1}\ha_{\ell} \right) + \frac{U}{2} \sum_{\ell=1}^{3} \ha^\dagger_\ell \ha^\dagger_\ell\ha_\ell\ha_\ell + \delta \left(\ha^{\dagger}_1\ha_1 - \ha^{\dagger}_3\ha_3 \right).
\end{align}
Again, moving to a frame rotating with angular frequency $\delta/\hbar$ using the unitary operator $\hat{R} = \mathrm{exp} \lbrace \di \delta t (\ha^{\dagger}_1\ha_1 - \ha^{\dagger}_3\ha_3 )/\hbar\rbrace$, we get the transformed Hamiltonian
\begin{align}
    \htH(t) &= J \sum_{\ell=1}^{2}\left(\de^{\di\delta t/\hbar} \, \ha^\dagger_{\ell}\ha_{\ell+1} - \de^{-\di\delta t/\hbar} \, \ha^\dagger_{\ell+1}\ha_{\ell} \right) + \frac{U}{2} \sum_{\ell=1}^{3} \ha^\dagger_\ell \ha^\dagger_\ell\ha_\ell\ha_\ell.
\end{align}
Using a similar procedure as for $L\!=2$ for $\delta \gg U,J$, one can write an effective Hamiltonian which till the second order of Magnus expansion is given as
\begin{align}
    \label{eq:h3well_tilted_2ndorder}
    \htH_\deff &\approx \left( \frac{U}{2} - \frac{J^2 U}{\delta^2}\right) \sum_{\ell=1}^{3} \ha^\dagger_\ell \ha^\dagger_\ell\ha_\ell\ha_\ell -  \frac{J^2 U}{\delta^2} \ha^\dagger_2 \ha^\dagger_2\ha_2\ha_2 + \frac{4 J^2 U}{\delta^2} \sum_{\ell=1}^{2} \ha^\dagger_{\ell}\ha_{\ell}\ha^{\dagger}_{\ell+1}\ha_{\ell+1} - \frac{J^2 U}{\delta^2} \left( \ha^\dagger_1 \ha^\dagger_3\ha_2\ha_2 + \mathrm{h.c.}\right) \nonumber\\
    &\quad + \frac{J^2}{\delta} \left( \ha^{\dagger}_1\ha_1 - \ha^{\dagger}_3\ha_3 \right).
\end{align}
As a matter of fact, the terms obtained here resemble the Hamiltonian describing rotationally invariant spin-1 gases \cite{pu_2000, evrard_2021},
\begin{align}
    \label{eq:hspin1}
    \frac{\hH_{\mathrm{S}=1}}{U_s} &= 2\hat{N} + \ha^\dagger_1 \ha^\dagger_1\ha_1\ha_1 + \ha^\dagger_3 \ha^\dagger_3\ha_3\ha_3 + 2\left(\ha^\dagger_1 \ha^\dagger_2\ha_1\ha_2 + \ha^\dagger_2 \ha^\dagger_3\ha_2\ha_3 - \ha^\dagger_1 \ha^\dagger_3\ha_1\ha_3  \right) + 2 \left(\ha^\dagger_1 \ha^\dagger_3\ha_2\ha_2 + \mathrm{h.c.} \right),
\end{align}
with $U_s$ the spin interaction energy. However, a one-to-one mapping between the two Hamiltonians does not exist, as $U,J$ and $\delta$ in Eq.~\eqref{eq:h3well_tilted_2ndorder} cannot be tuned to match the corresponding terms in Eq.~\eqref{eq:hspin1}; in this sense, the 3-mode tilted Bose-Hubbard model is not a direct analogue of spin-1 Bose gases.

\section{D.~ Quantum optimal control of $L$-tuple Fock states}
\label{app:qoc}
Given a dynamical system described by the state vector $\vec{x}(t)$ ($0\leq t\leq t_\df$), and  whose evolution is parameterized by control parameters $\vec{u}(t)$, optimal control~\cite{pontryagin_1987, bellman_1966, sussmann_1997, liberzon_book_2012} is the mathematical formalism that allows to determine $\vec{u}(t)$ in order to \textit{optimally} control $\vec{x}(t)$. In the framework of optimal control, \textit{optimality} is captured by a cost function $\mathcal{C}$ to be minimized. By virtue of the Pontryagin Maximum Principle, the optimal control $\vec{u}(t)$ is the one that maximizes the Pontryagin Hamiltonian $H_\dP(t), \; \forall t \in [0,t_\df]$ (see below, and Refs.~\cite{pontryagin_1987, bellman_1966, liberzon_book_2012, ansel_2024}). Quantum optimal control~\cite{glaser_2015,boscain_2021,koch_2022,ansel_2024} is the application of optimal control theory to quantum systems, with typically $\vec{x}(t) \rightarrow \ket{\psi(t)}$. 

\subsection{a.~ Implementation}
In this work, we use the energy bias $\delta(t)$ between the $L$-modes of a Bose-Hubbard system as our control parameter to optimally steer the ground state of the system (taken at $\delta=0$ with $U/J$ fixed) towards the $L$-tuple Fock state $\ket{N/L}$ in a fixed time $t_\df$. We do not place any constraint on $\delta(t)$. The cost function that we minimize is $\mathcal{C}=1-\mathcal{F}$, with $\mathcal{F} = |\braket{N/L}{\psi(t_\df)}|^2$, leading to the following definition of the Pontryagin Hamiltonian~\cite{liberzon_book_2012,dupont_2021,dupont_thesis_2022}:
\begin{equation}
    \label{eq:pontryagin}
    H_\dP(t) = \Re{\bra{\chi(t)}\dfrac{\partial}{\partial t}\ket{\psi(t)}} = \dfrac{1}{\hbar}\Im{\bra{\chi(t)}\hH(\delta(t))\ket{\psi(t)}},
\end{equation}
using Schr\"odinger equation with the Hamiltonian $\hH$ given in Eqs.~\eqref{eq:Hbjj_2well_tilt} and \eqref{eq:Hbjj_3well_tilt} for $L\!=2$ and 3 respectively. In Eq.~\eqref{eq:pontryagin}, $\ket{\psi(t)}$ is state during its preparation and $\ket{\chi(t)}$ is a non-normalized adjoint state, also solution of Schr\"odinger equation and defined by the final condition:
\begin{equation}
    \label{eq:adjoint_state}
    \braket{u_n}{\chi(t_\df)} = \dfrac{\partial\mathcal{F}}{\partial \left( \braket{\psi(t_\df)}{u_n} \right)},
\end{equation}
in any basis $\left\lbrace\ket{u_n}\right\rbrace$, giving $\ket{\chi(t_\df)} = \braket{N/L}{\psi(t_\df)}\ket{N/L}$.
Given an non-optimal control $\delta(t)$, the first-order correction $\Delta \delta(t)$ required to increase $H_\dP$ (and thus $\mathcal{F}$) is proportional to the derivative of $H_\dP$ with respect to the control parameter~\cite{boscain_2021, dupont_2021, ansel_2024}, that is
\begin{equation}
    \Delta \delta(t) = \dfrac{\epsilon}{\hbar} \Im{\bra{\chi(t)}\dfrac{\partial \hH(\delta(t))}{\partial(\delta(t))}\ket{\psi(t)}},
\end{equation}
with $\epsilon$ a small adjustable line-search parameter~\cite{boscain_2021,ansel_2024}. We iteratively correct $\delta(t) \rightarrow \delta(t) + \Delta \delta(t)$ until we reach the desired fidelity, following the algorithm given in Refs.~\cite{boscain_2021,dupont_2021,ansel_2024}.

As discussed in the main text, working around a large average tilt allows to restrict the quantum dynamics of the evolved state to a narrow subset of Hilbert space. We use it to our advantage by first computing an approach control field in a Hilbert space strongly truncated ``around'' $\ket{N/L}$~\footnote{For instance with $L=2$ and $N=128$, the first optimization was performed in a Hilbert space of dimension 19, only considering the basis Fock states $\ket{n_1}$ with $55\leq n_1 \leq 73$.}. We then use this first control field as a strong guess for a second optimization in the complete Hilbert space.

\subsection{b.~ Details of the controls featured in the main text}
\begin{figure}[t]
\begin{center}
\includegraphics[scale=1]{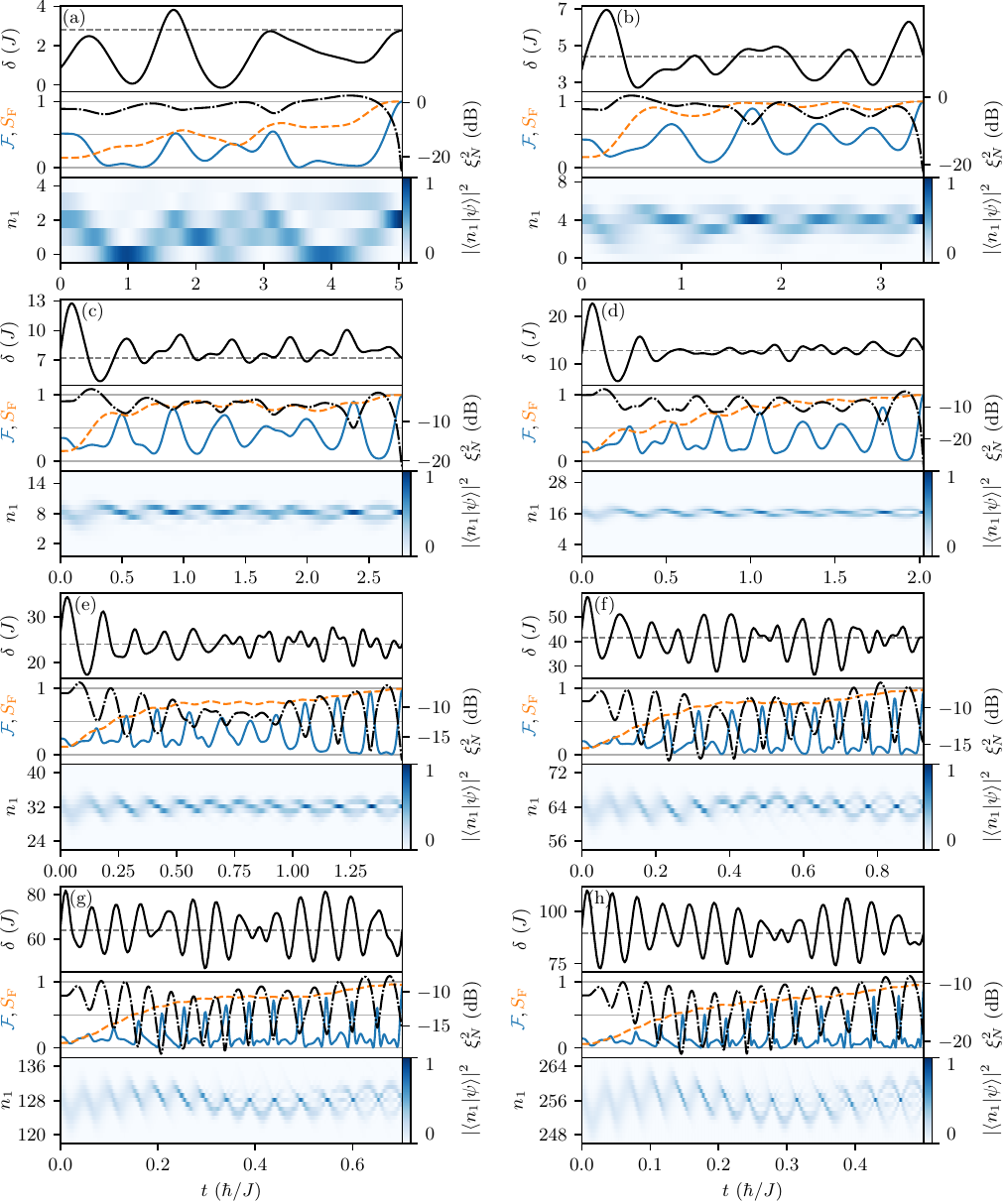}
\caption{
\textbf{Details of twin Fock state preparation by optimal control.} 
\textbf{(a)}~Top row: Optimal control tilt as a function of time (solid black line) to drive the ground state of the unbiased system with $U/J=1$ and $N=4$ towards the twin Fock state $\ket{N/2}$, and initial guess $\delta(t)=\delta_0$ (dashed grey line). Middle row: Fidelity $\mathcal{F}$ to $\ket{N/2}$ (blue solid line), fragmentation entropy $S_\dF$ (divided by $\ln(2)$, orange dashed) and number squeezing parameter $\xi^2_N$ (dB scale, black dash-dotted) as a function of time. Bottom row: Fock space projection during the preparation. \textbf{(b-h)}~Same as (a) for $N=8$, 16, 32, 64, 128, 256, 512 respectively. For (e-h), the displayed evolution in Fock space is cropped over $n_1=N/2\pm10$ (bottom rows).
}
\label{fig:qoc_details}
\end{center}
\end{figure}

We detail in Fig.~\ref{fig:qoc_details} the twin Fock state preparations presented in Table~1 of the main text, i.e. as a function of $N$. For each number of bosons considered, we show the optimal-control tilt $\delta$, the three metrics of state preparation considered in the main text (namely the fidelity $\mathcal{F}$ to the twin Fock state, the fragmentation entropy $S_\dF$ and the number squeezing parameter $\xi^2_N$), and the projection over the Fock basis as a function of the preparation time. In particular, panel (d), with $N=32$, corresponds to Fig.~2 of the main text.

\subsection{c.~ Robustness to number fluctuations}
\begin{figure}[t]
\begin{center}
\includegraphics[scale=1]{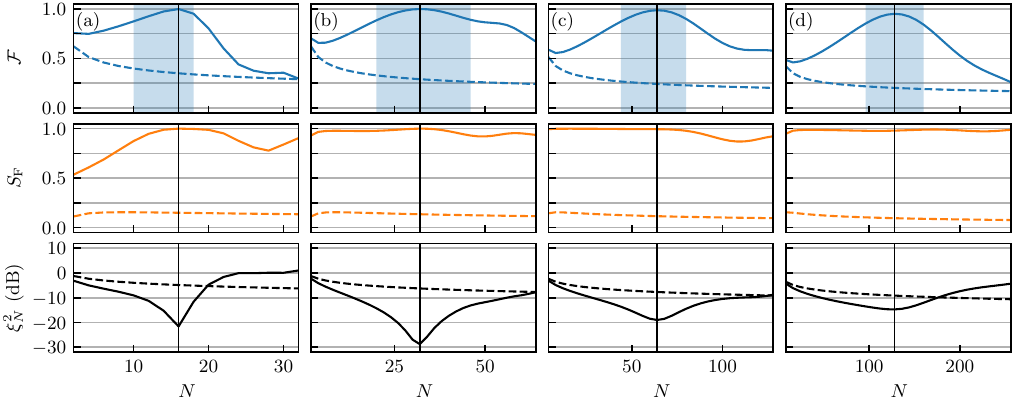}
\caption{
\textbf{Robustness of the optimal control preparations to number fluctuations.} Fidelity (top row), fragmentation entropy (divided by $\ln(2)$, middle row) and number squeezing (bottom row) as a function of $N$. Columns (a-d) correspond to control fields optimized for $N_0=16$ , 32, 64, 128 respectively (marked by the vertical black lines).
Solid lines shows the three metrics for the optimized states and dashed lines for the initial state (ground state at $U/J=1$ and $\delta=0$). 
The blue areas mark the intervals around $N_0$ over which the fidelity stays above 90\% of its maximum.
The corresponding control fields are plotted in Fig.~\ref{fig:qoc_details}(c-f).
}
\label{fig:qoc_robustness}
\end{center}
\end{figure}

To demonstrate the significant robustness of our optimal-control scheme to number fluctuations, which results from inducing fragmentation around a large tilt (see main text), we now apply the control fields that were optimized for a given $N_0$ to other settings with $N \ne N_0$. Figure~\ref{fig:qoc_robustness} shows those results for $N_0=16$, 32, 64 and 128, corresponding to the controls of Fig.~\ref{fig:qoc_details}(c-f). The fidelity remains above 90\% of its maximum over the large interval $N\in[10,18]$ for $N_0=16$, $N\in[20,46]$ for $N_0=32$, $N\in[44,80]$ for $N_0=64$ and  $N\in[96,160]$ for $N_0=128$. Figure~\ref{fig:qoc_robustness} further illustrates that the fidelity is a fragile metric compared to the fragmentation entropy, which remains very high around $N_0$, furthermore over a larger interval as $N$ increases.

We mention that dedicated approaches exist to enhance robustness in optimal control theory (see~\cite{ansel_2024} and references therein), however these go beyond the scope of this study.
    
\subsection{d.~ Quantum optimal preparations of triple Fock states}
\begin{figure}[b]
\begin{center}
\includegraphics[scale=1]{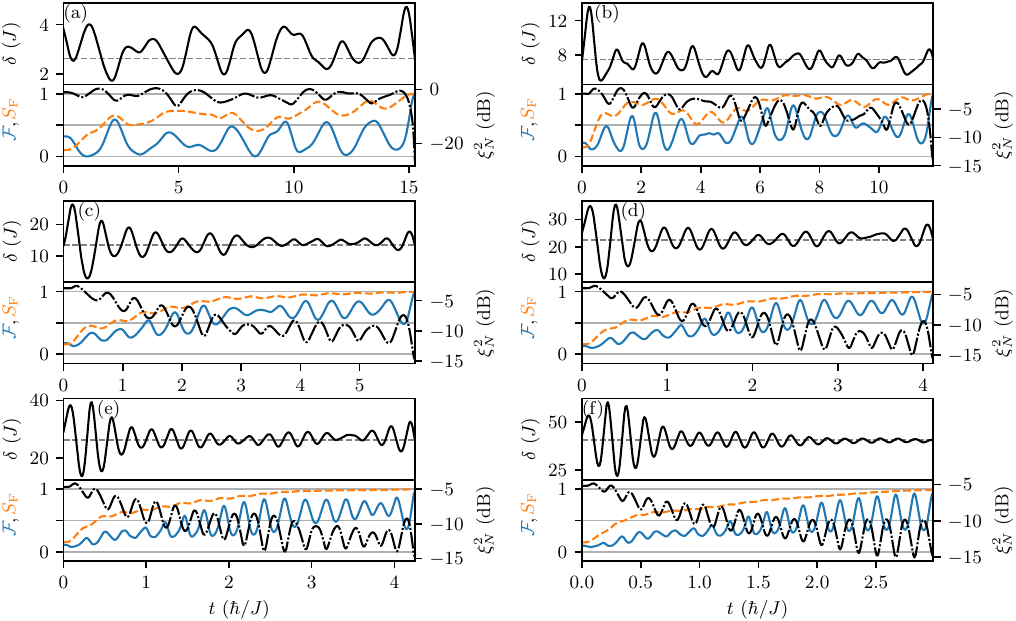}
\caption{
\textbf{Triple Fock state preparations by optimal control.}
\textbf{(a)}~Top row: Optimal control tilt as a function of time (solid black line) to drive the ground state of the unbiased system with $U/J=1$ and $N=3$ towards the triple Fock state $\ket{N/3}$, and initial guess $\delta(t)=\delta_0$ (dashed grey line). Bottom row: Fidelity $\mathcal{F}$ to $\ket{N/3}$ (blue solid line), fragmentation entropy $S_\dF$ (divided by $\ln(3)$, orange dashed) and number squeezing parameter $\xi^2_N$ (dB scale, black dash-dotted) as a function of time. \textbf{(b-f)}~Same as (a) for $N=6$, 9, 15, 21, 27 respectively. See Table~\ref{tab2} for details.
}
\label{fig:qoc_3well}
\end{center}
\end{figure}

Figure~\ref{fig:qoc_3well} shows results of triple Fock state preparation by optimal control for $N=3$, 6, 9, 15, 21 and 27 bosons. Our approach is the same as for $L\!=\!2$ (see main text and previous sections). Details of the preparations are given in Table~\ref{tab2}.

\newcommand\cwc{0.08}
\newcommand\cwd{0.0475}
\begin{table}[t]
\begin{center}
\begin{tabular}{
|>{\centering}p{\cwc\linewidth}|
>{\centering}p{\cwd\linewidth}
>{\centering}p{\cwd\linewidth}
>{\centering}p{\cwd\linewidth}
>{\centering}p{\cwd\linewidth}
>{\centering}p{\cwd\linewidth}
>{\centering\arraybackslash}p{\cwd\linewidth}|
}
\hhline{=======}
$N$ & 3 & 6 & 9 & 15 & 21 & 27 \\ \hline
$\delta_0 (J)$ & 2.63 & 7.50 & 13.5 & 22.5 & 26.3 & 40.5 \\
$t_\df\,(\hbar/J)$ & 15.3 & 11.8 & 5.93 & 4.12 & 4.25 & 2.99 \\
$\mathcal{F}$ & 0.999 & 0.990 & 0.980 & 0.971 & 0.962 & 0.937 \\
$S_\dF/\ln(3)$ & \multicolumn{2}{c}{$>0.999$} & 0.996 & 0.997 & 0.999 & 0.987 \\
\multirow{2}{*}{
\hspace{-2mm}$\begin{array}{c} \xi^2_N\\ \text{(dB)}\end{array}\hspace{-2mm}\left(\hspace{-1mm}\begin{array}{l} \psi_\df \\ \phi_0\end{array}\hspace{-1mm}\right)$
} & -26.9 & 14.4 & -14.8 & -15.7 & -14.9 & -14.9 \\
& -0.994 & -2.21 & -2.97 & -3.99 & -4.68 & -5.20 \\
\hhline{=======}
\end{tabular}
\caption{
\textbf{Triple Fock state preparations by optimal control} ($U/J=1$) for increasing number of bosons $N$, with guess tilt $\delta_0$, control duration $t_\df$, final fidelity $\mathcal{F}$ to $\ket{N/3}$, fragmentation entropy $S_\dF$ (divided by $\ln(3)$) and number squeezing parameter of prepared and initial states ($\ket{\psi_\df}$ and $\ket{\phi_0}$ resp.).
}
\label{tab2}
\end{center}
\end{table}

\section{E.~ Number fluctuations of reference}
\label{app:varJz2}
We derive, for $L\!=\!2$ and 3, the uncorrelated number fluctuations $\Delta \hJ^2_{z,\dref}$ used as the reference value when computing the number squeezing parameter $\xi^2_N(\psi) = \Delta \hJ^2_{z,\psi}/\Delta \hJ^2_{z,\dref}$. In each case considered, the reference state that is taken corresponds to the ground state of the hopping Hamiltonian with $N$ bosons.

\subsection{a.~ Uncorrelated number fluctuations for $L\!=\!2$} 
The reference state is the coherent spin state given in Eq.~\eqref{eq:gs_jx} for $n=0$, explicitly
\begin{equation}
    \label{eq:gs_jx2}
    \ket{\phi^{(0)}_0} = \dfrac{\left( \ha^\dagger_1 + \ha^\dagger_2 \right)^N}{2^{N/2}\sqrt{N!}}\ket{0,0}.
\end{equation}
Number fluctuations for that state are given by the variance on $\hJ_z$: $\Delta \hJ^2_{z,\dref} = \langle \hJ_z^2 \rangle_\dref - \langle\hJ_z\rangle^2_\dref$, with $\hJ_z = (\ha^\dagger_1\ha_1-\ha^\dagger_2\ha_2)/2$. As the reference state lies symmetrically on the equator of the Bloch sphere, the first moment of $\hJ_z$ is zero:
\begin{align}
    \langle\hJ_z\rangle_\dref &=  \dfrac{1}{2^{N+1}N!}\bra{0,0}\left(\ha_1+\ha_2\right)^N\left(\ha^\dagger_1\ha_1-\ha^\dagger_2\ha_2\right)\left(\ha^\dagger_1+\ha^\dagger_2\right)^N\ket{0,0} \notag\\
    &= \dfrac{1}{2^{N+1}N!}\left(\sum_{n=0}^N \binom{N}{n} \sqrt{(N-n)!n!}\bra{N-n,n}\right)\left(\ha^\dagger_1\ha_1-\ha^\dagger_2\ha_2\right)\left(\sum_{m=0}^N \binom{N}{m} \sqrt{(N-m)!m!}\ket{N-m,m}\right) \notag\\
    &=\dfrac{1}{2^{N+1}}\sum_{n=0}^N \binom{N}{n}(N-2n) = 0,
\end{align}
where the last equality is evaluated using the following identities on the sum of the binomial coefficients:
\begin{equation}
    \label{eq:binomial_prop1}
    \sum_{n=0}^N\binom{N}{n} = 2^N \qquad \text{and} \qquad \sum_{n=0}^Nn\binom{N}{n} = 2^{N-1}N.
\end{equation}
Number fluctuations are thus given by the second moment of $\hJ_z$, that is
\begin{equation}
    \label{eq:2well_numb_fluct}
    \Delta \hJ^2_{z,\dref} = \langle \hJ_z^2 \rangle_\dref
    = \dfrac{1}{2^{N+2}}\sum_{n=0}^N \binom{N}{n}(N-2n)^2 = \dfrac{N}{4},
\end{equation}
using Eqs.~\ref{eq:binomial_prop1} and
\begin{equation}
    \label{eq:binomial_prop2}
    \sum_{n=0}^N n^2\binom{N}{n} = 2^{N-2}N(N+1).
\end{equation}
The result in Eq.~\eqref{eq:2well_numb_fluct} is well known, see e.g. Refs.~\cite{kitagawa_1993,spekkens_1999}

\subsection{b.~ Uncorrelated number fluctuations for $L\!=\!3$}
For $L=3$, we evaluate the number fluctuations \textit{partially}, via the variance of the population imbalance between modes 1 and 3: $\hJ_z = \ha^\dagger_1\ha_1-\ha^\dagger_3\ha_3$~\cite{mueller_2006, pezze_2018}. The hopping Hamiltonian (without coupling between modes 1 and 3) is
\begin{align}
    \label{eq:Htun_3well]}
    \hH &= -J \left( \ha^\dagger_1\ha_2 + \ha^\dagger_2\ha_3 + \ha^\dagger_2\ha_1 + \ha^\dagger_3\ha_2 \right) \notag\\
    &= -\sqrt{2}J \left( \ha^\dagger_+ \ha_+ - \ha^\dagger_-\ha_- \right),
\end{align}
with $\ha_\pm = (\ha_1 \pm \sqrt{2} \ha_2 + \ha_3)/2$. The reference state for number fluctuations is the ground state of Hamiltonian~\eqref{eq:Htun_3well]}, namely
\begin{align}
    \label{eq:gs_jx_3chain}
    \ket{\phi^{(0)}} &= \dfrac{\left( \ha^\dagger_1 \pm \sqrt{2} \ha^\dagger_2 + \ha^\dagger_3 \right)^N}{2^N\sqrt{N!}}\ket{0,0,0} \notag\\
    &= \dfrac{1}{2^N\sqrt{N!}}\sum_{n=0}^N\binom{N}{n}2^{(N-n)/2}\sqrt{(N-n)!}\sum_{r=0}^n\binom{n}{r}\sqrt{(n-r)!r!}\ket{n-r,N-n,r},
\end{align}
which is not a coherent spin state in the absence of coupling between modes 1 and 3. For that state (having equal populations in modes 1 and 3), the first moment of $\hJ_z$ is zero:
\begin{equation}
    \langle\hJ_z\rangle_\dref = \dfrac{1}{2^{N}}\sum_{n=0}^N\binom{N}{n}\dfrac{1}{2^n}\sum_{r=0}^n\binom{n}{r}(n-2r) = 0,
\end{equation}
as the second sum is null from Eqs.~\eqref{eq:binomial_prop1}. The variance of $\hJ_z$ is thus
\begin{equation}
    \Delta \hJ^2_{z,\dref} = \langle \hJ_z^2 \rangle_\dref
    = \dfrac{1}{2^{N}}\sum_{n=0}^N\binom{N}{n}\dfrac{1}{2^n}\sum_{r=0}^n\binom{n}{r}(n-2r)^2 = \dfrac{N}{2},
\end{equation}
using Eqs.~\eqref{eq:binomial_prop1} and~\eqref{eq:binomial_prop2}.

For completeness, we give that the uncorrelated number fluctuations for $L\!=\!3$ \textit{with} coupling between modes 1 and 3 (i.e. considering a coherent spin state, e.g. the ground state of a ring of three spatial sites) are given by $\Delta \hJ^2_{z,\dref} = 2N/3$.

\end{document}